\def\ignorethis#1{{}} 
\font\ztitle=cmr10 scaled\magstep4
\baselineskip=\normalbaselineskip
\def\newpage{\vfill\eject}

\font\zmsec=cmbx10 scaled 1200
\def\zmsecspc{{\vskip16pt}}
\def\zmtsecspc{{\bigskip}}

\newcount\equationnumber
\newcount\sectionnumber
\sectionnumber=1
\def\zen{\the\sectionnumber.\the\equationnumber}
\def\zp{
   \global\advance\equationnumber by 1 
   (\zen)}
\def\zpa#1{
   \global\advance\equationnumber by 1 
   (\zen{{\rm #1}})}
\def\zps#1{(\zen{{\rm #1}})}
\def\advsecnum{
   \equationnumber = 0
   \global\advance\sectionnumber by 1}

\def\zsecone{1}
\def\zsectwo{2}
\def\zsecthree{3}
\def\zsecfour{4}

\def\a{\alpha}

\def\bl{\left(}
\def\br{\right)}
\def\bt{{\tilde b}}

\def\d#1{\partial_{#1}}
\def\dg{\dagger}
\def\dt{\delta}
\def\dd{\Delta}

\def\ep{\epsilon}

\def\Eimp{E_{\rm imp}}

\def\fa{{1\over\sqrt{2}}}

\def\gg{\Gamma}

\def\zix{{\int_{-\infty}^0 dx\,}}
\def\zit{{\int_{-\infty}^\infty dt\,}}
\def\zik{{\int_{-\infty}^\infty {dk\over 2\pi}\,}}

\def\zu{{\uparrow}}
\def\zd{{\downarrow}}

\def\p{\psi}
\def\pb{{\overline\psi}}

\def\rb#1{{ |#1\rangle}}

\def\t{\theta}
\def\s{\sigma}

\def\w{\omega}

\def\wt{{\tilde\omega}}

\def\vk{{\vec k}}
\def\vx{{\vec x}}
\def\vf{{V\over\sqrt{2}}}

\def\abrikosov{1}
\def\kondo{2}
\def\anderson{3}
\def\vonsovsky{4}
\def\hartmann{5}
\def\satori{6}
\def\jarrell{7}
\def\borkowski{8}
\def\rupasov{9}
\def\zvyagin{10}
\def\yazdani{11}
\def\bassilec{12}
\def\wiegmann{13}
\def\machida{14}
\def\tinkham{15}
\def\ghoshal{16}
\def\fradkin{17}

\def\bsengyintZ{3.15}



\headline={\hfil CLNS 98/1593;\ cond-mat/9811312}
\footline={7/99\hfil}

\topskip=1in
\centerline{{\ztitle Bound States for 
 a Magnetic  Impurity}}
\smallskip
\centerline{{\ztitle in a Superconductor}}
\vskip1in
\centerline{Zorawar S. Bassi$^\dg$}
\centerline{Andr\'e LeClair}
\bigskip
\centerline{Newman Laboratory}
\centerline{Cornell University}
\centerline{Ithaca, NY 14853, USA}
\vskip0.6in
\noindent
We discuss a solvable model describing an Anderson like 
impurity in a BCS superconductor.  The
model can be mapped onto an  Ising field theory in a boundary
magnetic field, with the Ising
fermions being the quasi-particles of the Bogoliubov transformation 
in BCS theory.  
The reflection S-matrix exhibits Andreev scattering 
and the existence of bound states
of the quasi-particles with the impurity 
lying inside the superconducting gap.

\smallskip
\noindent
\vskip3.95in
\noindent
$^\dg$ zorawar@mail.lns.cornell.edu

\newpage
\advance\pageno by -1
\footline={\hfil\folio\hfil}
\headline={\hfil}

 \topskip=10pt

\noindent
{\zmsec Introduction}\par
\nobreak
\zmtsecspc

The problem of magnetic impurities in superconductors is an 
old yet still very interesting problem.  
Starting with the
work of Abrikosov and Gor'kov [\abrikosov], numerous theoretical studies have
addressed this problem.  The models most frequently used to 
describe the magnetic impurities are the Kondo
(s-d exchange) model [\kondo] and some form of the Anderson model [\anderson].
Most earlier works are based on a perturbative
approach.  A review of the earlier works
and various references can be found in [\vonsovsky].  

An important effect due to the impurities involves the existence of localized
excited states within the superconducting energy gap.  The results of
M{\"u}ller-Hartmann and Zittartz 
on gap states due to a Kondo impurity are well known [\hartmann].  
More recent studies of the gap states include a numerical
renormalization group analysis for a Kondo impurity [\satori], 
quantum Monte Carlo
techniques applied to the symmetric Anderson model [\jarrell], 
large-N slave boson
methods for an N-fold degenerate Anderson impurity [\borkowski] and 
a Bethe ansatz
analysis for an Anderson impurity by Rupasov [\rupasov].  
The closely related problem of an Anderson like impurity in 
a Hubbard chain, which has a spin gap, was studied in [\zvyagin].    
Recent experimental work on single impurities 
was reported in [\yazdani].  

In this paper we consider a rather simplified version of the
problem of  a magnetic impurity in
a BCS superconductor.  
The virtues of our model are that it is solvable by
elementary methods, and  also that it 
reveals some of the expected features of more
complicated models.  Our solution represents a self-consistent
treatment of the BCS pair potential and a fully quantum impurity.
We came to consider this model after a detailed
study of the more difficult theory corresponding to a 
bulk mass term in the usual Kondo model [\bassilec].  
Here, the interaction with the impurity is taken to be
of the Anderson form.  However unlike the Anderson model, the impurity
`d' orbital is considered to have no spin (or orbital) degeneracy, hence
it is described by a single fermion
 operator $d$ and its adjoint $d^\dg$, with
no spin index.  Since the conduction electrons have spin, the
impurity interaction for this model is intermediate to that of the Anderson
model and the resonant level model [\wiegmann] at 
the Toulouse point.  We study the 
model by mapping it onto a boundary field theory (BFT) and solving for the
reflection S-matrix (i.e. boundary scattering matrix).  The BFT consists
of two decoupled Ising models with boundary magnetic fields.    The Ising
fermions can be associated with the quasi-particles resulting from a
Bogoliubov transformation.  {}From this mapping one sees that whereas
a Kondo impurity has no non-trivial flow in boundary conditions, 
an Anderson impurity does have a flow between free and fixed boundary
conditions.  A similar model, where the localized `d' electrons do have
a spin degeneracy, i.e. the Anderson model with no Coulomb repulsion ($U=0$),
was studied some time ago in [\machida] using the more conventional 
Green's functions approach.

Our solution exhibits a single-impurity version of Andreev scattering.
Andreev scattering occurs at the boundary between a superconductor
and an ordinary metal, wherein quasi-particles are reflected into
quasi-holes.  For a single magnetic impurity, superconductivity 
can be destroyed locally in the vicinity of the impurity by the 
magnetic field of the impurity, in analogy with the Meissner effect.  
Thus one is led to expect that the impurity reflects quasi-particles
into quasi-holes, and our solution confirms this.  

It is well-known that 
magnetic impurities compete with superconductivity in a manner analogous to
the  
Meissner effect:  
if the concentration of impurities is large, enough bound states can
open up in the gap that the gap disappears, destroying superconductivity.
For a single impurity the competition is between the formation of
Cooper pairs and the binding of electrons to the impurity, and is
manifested in the properties of the bound states localized at the
impurity.  
In our model, from the poles in the reflection amplitude 
for each Bogoliubov quasi-particle
we can determine the bound states of the  
quasi-particle with the impurity.  
Simple energetics determines whether the bound state is stable. 
The impurity has a magnetic energy scale $E_{\rm imp}$.
When the gap $\Delta$ is zero, 
this energy scale appears as a resonance in the reflection S-matrix,
and corresponds to an unstable bound state of the original electrons
with the impurity.
(In the doubled boundary Ising description, this resonance corresponds
to a splitting in the degeneracy of the two Ising ground states induced
by the boundary magnetic field.) 
When the gap is non-zero, the state of energy $\Eimp$
 can become stable if it is forbidden to decay 
into the bulk superconductor because of the gap; 
this occurs   if $\Delta > \Eimp$.    
This bound state has an excitation 
energy, $E_b$, less than the superconducting gap $\Delta$. 
The bound state energy is a function of $\Eimp$ and $\Delta$,
and is given in (\bsengyintZ) below.
We find that the  bound states in the gap disappear 
when the strength of the impurity is large enough, i.e. when
$\Eimp \geq \Delta$.
In this case the `Meissner effect' is reversed and superconductivity
is favored: 
the quasi-particle-impurity bound state
becomes unstable and it is more energetically favorable to form 
Cooper pairs. 


\zmsecspc
\noindent
{\zmsec \zsecone. The Model}\par
\nobreak
\zmtsecspc

The model we use to describe an impurity in a superconductor consists of
the (mean-field) BCS Hamiltonian and an Anderson like interaction,
$$ H = H_{\rm BCS} + H_I \eqno\zp $$
\edef\horigZ{\zen}%
$$ H_{\rm BCS} = \sum_{\vk \s} \xi_\vk c_{\vk \s}^\dg c_{\vk \s}
- \dd \sum_\vk \bl c_{\vk \zu}^\dg c^\dg_{-\vk \zd} + 
c_{-\vk \zd} c_{\vk \zu} \br \eqno\zp $$
$$ H_I = \sum_{\vk \s} V_\vk\bl c^\dg_{\vk \s} d + d^\dg c_{\vk\s}
\br, \eqno\zp $$
where $c_{\vk \s}^\dg$ are conduction electron creation operators of
momentum $\vk$ and spin $\s$ ($\zu$ up and $\zd$ down), and
$\dd$ is the energy gap.  The impurity is at the origin in 
three spacial dimensions.      The
operators $d$ and $d^\dg $ describe a
spinless fermion localized on the impurity, satisfying  
$$  \{ d , d^\dg \} = 1, \quad d^2 = {d^\dg}^2 = 0. \eqno\zp $$
These operators act on a two-dimensional Hilbert space 
$\rb{0}_d$, $\rb{1}_d $ where $\rb{1}_d = d^\dg \rb{0}_d $. 
In the unperturbed theory these two states are taken to be degenerate
in energy. 
Since the impurity level has no spin degeneracy,
there is no Coulomb repulsion term as in the Anderson model.  
The energy $\xi_\vk$ is measured relative to the fermi energy
$\ep_F$, $\xi_\vk = \ep_\vk - \ep_F$, where $\ep_\vk$ is the electron kinetic
energy.  In the impurity interaction $H_I$, $V_\vk$ is the hybridization
matrix element between the band states and the impurity states.

We can diagonalize the BCS part by a 
Bogoliubov unitary transformation [\tinkham] 
$$ c_{\vk\zu} = u_\vk \bt_{\vk 1} + \a_\vk v_\vk \bt^\dg_{\vk 2},\quad
c_{-\vk \zd}^\dg =  u_\vk \bt_{\vk 2}^\dg -\a_\vk v_\vk \bt_{\vk 1},
 \eqno\zp $$
with
$$ u_\vk = \fa \sqrt{1 + \a_\vk {\xi_\vk\over \w_\vk}}, \quad
v_\vk = \fa \sqrt{1 - \a_\vk {\xi_\vk \over \w_\vk}} \eqno\zp $$
\edef\coefbogZ{\zen}%
$$ \a_\vk = {\rm sgn}(|\vk| - k_F) = \cases{ 1, & if $|\vk| > k_F$; \cr
-1, & if $|\vk| < k_F$, \cr } \eqno\zp $$
and
$$ \w_\vk = \sqrt{ \xi_\vk^2 + \dd^2 }. \eqno\zp $$
In terms of the quasi-particle operators $( \bt_{\vk j}^\dg,\bt_{\vk j})$,
which are often referred to as Bogoliubons, $H_{\rm BCS}$ becomes
$$ H_{\rm BCS} = \sum_{\vk j} \wt_\vk \bt_{\vk j}^\dg \bt_{\vk j}, \eqno\zp $$
\edef\hbcsbtZ{\zen}%
where
$$ \wt_\vk = \cases{ -\w_\vk, & if $|\vk|  < k_F$; \cr \w_\vk, & if 
$|\vk| >k_F$.\cr } \eqno\zp $$
The ground state of (\hbcsbtZ), $\rb{0}_b$, is obtained by filling all
levels with $|\vk|< k_F$.
Rather than working with the $\bt$'s, it is more convenient to 
introduce the following `modified' quasi-particle operators
$$ b_{\vk 1} = \cases{ \bt_{\vk 1}, & if $|\vk|> k_F$; \cr \bt_{\vk 2}, & if
$|\vk| < k_F$, \cr} \qquad
b_{\vk 2} = \cases{ \bt_{\vk 2}, & if $|\vk|> k_F$; \cr \bt_{\vk 1}, & if
$|\vk| < k_F$. \cr} \eqno\zp $$
These operators satisfy canonical commutation relations and will simply be
referred to as the quasi-particles of the system.  The BCS Hamiltonian
maintains its diagonal form (\hbcsbtZ) when expressed in this new basis. 

The Bogoliubov transformation applied to the interaction gives
$$ H_I = {V\over\sqrt{2}}\sum_\vk 
\left [ (b_{\vk 1}^\dg - b_{\vk 1} ) (d+ d^\dg) +
(b_{\vk 2}^\dg + b_{\vk 2} ) (d- d^\dg) \right ], \eqno\zp $$
\edef\hintbtZ{\zen}%
where we have set $\vk = \vk_F$ in $V_\vk$, $V= V_{k_F}$, and in
the coefficients of the unitary transformation (\coefbogZ).  In (\hintbtZ)
we have kept the terms $b^\dg_{\vk j} d^\dg$ and $b_{\vk j} d$, which were 
omitted in the `rotating wave approximation' made in [\rupasov].

Assuming the impurity to be point-like with only s-wave scattering, we
expand the operators
$( b_{\vk j}^\dg,b_{\vk j})$ in spherical harmonics and retain only
the angular momentum $l=m=0$ terms (the $l\neq 0$ terms do not couple to the
impurity).  This gives the one dimensional Hamiltonian (sum over $j$)
$$ H = \int_{-\infty}^\infty 
{dk\over 2\pi}  \bl \wt(k) b^\dg_j(k) b_j(k) +
{V\over\sqrt{2}} \left [ (b_1^\dg(k) - b_1(k) ) (d+ d^\dg) +
(b_2^\dg(k) + b_2(k) ) (d- d^\dg) \right ] \br, \eqno\zp $$
where
$$ \wt(k) = \cases{ -\w(k), & if $k  < 0$; \cr \w(k), & if 
$k > 0$,\cr } \quad \w(k) = \sqrt{ k^2 + \dd^2}. \eqno\zp $$
We have linearized $\xi_\vk$ about the fermi surface, $\xi_\vk = |\vk|-k_F$,
setting $v_F = 1$, and defined $k=|\vk| - k_F$ to be the momentum relative
to the fermi surface.  The operators $b_j^\dg(k)$ create quasi-particles
of momentum $k+k_F$ and satisfy 
$$ \{ b_i(k),b_j^\dg(k^\prime) \} = 2 \pi \dt_{i j} \dt(k-k^\prime). 
\eqno\zp$$
Note that $V$ has dimension $\sqrt{{\rm energy}}$;  below we show that
the scale $\Eimp$ introduced in the introduction is simply $V^2$.   

A final transformation
$ i b_1(k) \rightarrow b_1(k)$ and  $-i b_1^\dg(k) \rightarrow b_1^\dg(k)$,
allows $H$ to be rewritten as
$$ H = H_1 + H_2 \eqno\zp $$
\edef\hsplitZ{\zen}%
$$ H_j = \int_{-\infty}^\infty 
{dk\over 2\pi} \, \bl \wt(k) b^\dg_j(k) b_j(k) +
i {V\over\sqrt{2}}\bl b_j^\dg(k) + b_j(k)\br a_j \br \quad (j=1,2), \eqno\zp $$
\edef\hfinZ{\zen}%
where
$$ a_1 = (d + d^\dg),\quad a_2 = -i(d-d^\dg). \eqno\zp $$
The operators $a_j$ anticommute with the $b$ operators and satisfy
$$ a_j^\dg = a_j, \quad a_j^2 = 1, \quad \{a_1, a_2\} = 0, \eqno\zp $$
implying
$ [H_1,H_2] = 0$.
Thus we only need to concentrate on one copy and henceforth drop 
the subscript.

\advsecnum

\zmsecspc
\noindent
{\zmsec \zsectwo. Mapping onto a Boundary Field Theory}\par
\nobreak
\zmtsecspc

In this section we map (\hfinZ) onto a boundary Ising field theory.
As in the usual Kondo model, s-wave scattering has reduced the 
problem to a one-dimensional problem on the half-line $r\geq 0$, where
$r$ is the spherical radial coordinate.  Henceforth, $r$ is designated 
as the spacial variable $-x$.  

The action for an Ising field theory on the half-line with a
magnetic field at the boundary is [\ghoshal]
$$ S = S_{\rm bulk} + S_{\rm a-free} + S_{\rm bc}+ S_{\rm int} \eqno\zp$$
\edef\stotalZ{\zen}%
$$ S_{\rm bulk} = {i\over 2} \zit\zix 
\bl \p(\d t - \d x)\p + \pb(\d t + \d x)\pb + 2\dd \p \pb \br \eqno\zp $$
$$ S_{\rm bc} = -{i\over 2} \zit \p\pb{\big |}_{x=0} , 
\qquad
S_{\rm a-free} = {i\over 2} \zit a(t)\d t a(t)
\eqno\zp $$
$$ S_{\rm int} = - i \vf \zit (\p + \pb) a(t){\big |}_{x=0}. \eqno\zp $$
The term $S_{\rm bulk}$ is simply a free Majorana action with $\dd$ being
the fermion mass, $S_{\rm bc}$ serves to enforce the free boundary condition 
$$ \p = \pb \quad {\rm at\ } x=0, \eqno\zp $$
for $V=0$, and $S_{\rm int}$ gives the interaction between the boundary spins
and the boundary magnetic field $V$.
The fermionic field $a(t)$, which anticommutes with $(\p, \pb)$, 
describes the ground state degeneracy due 
to the two different expectation values of the spin field and satisfies 
$a^2 =1$.  Thus $S_{\rm a-free}$ is the kinetic term for $a(t)$.
The bulk equations of motion are
$$ (\d t - \d x)\p + \dd \pb = 0, \quad (\d t + \d x)\pb - \dd \p =0.
\eqno\zp $$
\edef\eomZ{\zen}%
Mode expansions satisfying (\eomZ) can be written as
$$ \left( \matrix{\p \cr \pb \cr} \right)
 = \int_{-\infty}^{\infty} {dk\over 2\pi}\ \left[
\left( \matrix{ u(k) \cr v(k) } \right)  A(k) e^{-i \vk\cdot\vx} 
+ \left( \matrix{ u^* (k) \cr v^* (k) \cr } \right) 
 A^\dg (k) e^{i \vk\cdot\vx} \right],
\eqno\zp $$
where
$$  \{A(k),A^\dg(k^\prime)\} = 2\pi \delta(k-k^\prime)\eqno\zp $$
$$ u(k) = e^{-i\pi/4} \sqrt{{\w(k) - k \over 2 \w(k)}},\quad
 v(k) = e^{i\pi/4} \sqrt{{\w(k) + k \over 2 \w(k)}},
 \eqno\zp $$
and 
$ \vk\cdot\vx = \w(k) t - k x$,  $\w (k) = \sqrt{k^2 
+ \Delta^2}$. 

The Ising Hamiltonian associated with (\stotalZ) takes the form
$$ H = \zik \bl \w(k) A^\dg(k) A(k) + i\vf \bl
g(k) A(k) + g^*(k) A^\dg(k) \br a \br, \eqno\zp$$
\edef\btothZ{\zen}%
where
$$ g(k) = u(k) + v(k). \eqno\zp$$
The second term in (\btothZ) gives the boundary interaction.
Comparing (\btothZ) with (\hfinZ), we can map each $H_j$ onto the Ising
Hamiltonian if we relate the $b$ operators to the $A$ operators as follows
$$ b(k) = g(k) A(k)\quad {\rm if \ }k>0 \eqno\zpa{a}$$
$$ b(k) = g^*(k) A^\dg(k)\quad 
{\rm if \ }k<0,\eqno\zps{b} $$
and identify $a_j$ with $a$.  The quasi-particles correspond to the Ising 
fermions and the impurity coupling $V$ plays the role of a 
boundary magnetic field.  Using
$$ g(k) g^*(k) = 1, \eqno\zp $$
it is easily shown that the $b$ operators satisfy canonical commutation
relations
$$ \{ b(k) ,b^\dg(k^\prime)\} = 2\pi \dt(k-k^\prime). \eqno\zp $$

\advsecnum

\zmsecspc
\noindent
{\zmsec \zsecthree. Quasi-Particle Bound States}\par
\nobreak
\zmtsecspc

Having mapped (\hfinZ) onto the BFT (\stotalZ), the impurity interaction
can then be described by the reflection S-matrix (boundary scattering matrix)
for the fermion operators.  We compute the reflection amplitude from the
action as discussed in [\ghoshal].   The boundary 
terms in the variation of
(\stotalZ) lead to the boundary equation of motion (at $x=0$)
$$ - \d t(\p - \pb) = V^2(\p + \pb). \eqno\zp $$
\edef\beomZ{\zen}%
Here we see the flow between free ($V=0$) and fixed ($V=\infty$) boundary
conditions.
Interpreting (\beomZ) to hold when acting on a formal boundary operator
$B$, we substitute the mode expansions and define the reflection S-matrix,
or reflection amplitude, $R(k)$, through
$$ A^\dg(k) B = R(k) A^\dg(-k) B. \eqno\zp $$
The reflection amplitude for the Ising model was computed in [\ghoshal].  
Let us  parametrize the energy and momentum with the rapidity $\t$ 
$$ \w(\t) = \dd \cosh\t, \quad k(\t) = \dd\sinh\t. \eqno\zp $$
Define
$ A(\t) = A(k)\sqrt{\w(k)}$, satisfying 
$ \{ A(\t), A^\dg(\t^\prime) \} = 2 \pi \dt(\t - \t^\prime)$.  
We find 
$$ A^\dg(\t) B = R(\t) A^\dg(-\t)B, \eqno\zp$$
where
$$ R(\t) =  -i \tanh\bigl(i{\pi\over 4} - {\t\over 2}\bigr)
R_b (\t ) , \qquad
 R_b (\t)  =  \left(
{ {i\sinh\t + (V^2/\dd - 1)}\over{i\sinh\t -(V^2/\dd - 1)} }
\right).
 \eqno\zp$$

The amplitude $R_b$ is the reflection amplitude for the 
Bogoliubov operators $b(k)$
$$ b^\dg(k>0) B = R_b(\t) b(-k) B,\qquad b(k<0) B = R_b(\t) b^\dg(-k) B.
\eqno\zp $$
\edef\rmbZ{\zen}%
Defining  the quasi-electron operators $e(k)$
for $k>0$ and quasi-hole operators $h(k)$ for $k<0$
$$ e(k) = b(k), ~~ {\rm for} ~ k>0; \qquad
 h(k) = b^\dg(k), ~~ {\rm for} ~ k<0,  \eqno\zp $$
then (\rmbZ) can be expressed as
$$ e^\dg(k) B = R_b(k) ~  h^\dg(-k) B,\qquad 
h^\dg(k) B = R_b(k) e^\dg(-k) B. \eqno\zp $$
\edef\qetqhsZ{\zen}%
The quasi-electrons  scatter into quasi-holes and vice-versa 
at the boundary.  This is a form of Andreev reflection, which usually occurs
at the boundary between a superconductor and a normal metal.  Here the
boundary is a single impenetrable impurity,
i.e. there is no transmission. For the original quasi-particles, the 
$\bt$'s, equation (\qetqhsZ) implies that quasi-electrons of one type, 
say $\bt^\dg_1(k)$, scatter into quasi-holes of the other type $\bt_2(-k)$.

In terms of the energy $E(\t) = \w(\t)$, $R_b$ can be written as
$$ R_b(E) = - {E^2 + (V^2)^2 - 2\dd V^2 \over (V^2)^2 - E^2 - 
2(V^2 -\dd)(\dd + i \sqrt{E^2 - \dd^2}) }.\eqno\zp $$
\edef\remfZ{\zen}%
For $\dd=0$, (\remfZ) becomes
$$ R_b^0(E) = - { \Eimp^2 + E^2 \over \Eimp^2 - E^2 - i\gg E}, \eqno\zp$$
where
$$ \Eimp = V^2, \quad \gg = 2 V^2.\eqno\zp $$
The energy $\Eimp$ is a `magnetic energy' scale associated with the magnetic 
impurity. This becomes clearer in the Ising picture where 
$V$ effectively acts as a local magnetic field.
Near $E\approx \Eimp$, $R_b^0(E)$ takes a Lorentzian form
$$ R_b^0(E) = -i { (\gg/2)^2 \over (\Eimp - E)^2 + (\gg/2)^2 }. \eqno\zp $$
\edef\rlorZ{\zen}%
{}From (\rlorZ) we see that when $\dd=0$, there is a resonance in the
reflection amplitude at $E=\Eimp$ with a width $\gg$.  
This resonance corresponds to an 
unstable electron-impurity state with an energy $\Eimp$
and decay rate $\Gamma$.  If  
the gap $\dd$ is non-zero, then this state can become
a stable bound state (see below).

As with the bulk S-matrix, poles of the reflection S-matrix provide 
information on bound states.  Poles in the physical strip 
$0\leq {\rm Im}\,\t \leq \pi$ can be indicative of either bulk bound states or
boundary bound states.  Since the bulk theory is a free fermion theory
with trivial scattering ($S=-1$), we only have to consider  the 
latter case.  The quasi-particle reflection amplitude is
$$ R_b(\t) = { {i\sinh\t + (\Eimp/\dd - 1)} \over
{i\sinh\t -(\Eimp/\dd - 1)} }.\eqno\zp$$
We will now express all quantities in terms of the 
physical energy scale $\Eimp$.
Writing $\t$ as $iu$, poles of $R_b(\t)$ occur when
$$ \sin u = 1 - {\Eimp\over \dd}. \eqno\zp $$
\edef\pcZ{\zen}%
Restricting $u$ to be in the physical strip, there is a single pole at
$u=u_b$ satisfying (\pcZ) provided $\Eimp \leq \dd$.
(Note that there is also a pole at ${\tilde u}_b = \pi - u_b$, corresponding
to scattering in the cross channel.  The results that follow also apply
to this pole.)  Thus a quasi-particle forms a bound state with the 
impurity if $\Eimp \leq \dd$.  This is the same state that appears as a 
resonance in (\rlorZ) for a vanishing gap;  it has become stable 
because its decay into the bulk is forbidden if $\Eimp < \Delta$.   
The excitation energy (above the ground state) of the bound state, $E_b$, is 
$$ E_b = \dd\cos u_b = \sqrt{ \dd \Eimp \bl 2 - {\Eimp\over \dd} \br}.
\eqno\zp $$
\edef\eeZ{\zen}%
For $\Eimp < \dd$, the excitation energy lies within the
superconducting gap ($E_b<\dd$) and the bound state is stable.   
When $\Eimp>\dd$, the excitation energy falls in the continuum and there is
no bound state.

If $\Eimp=0$, the pole
occurs at $i\pi/2$, implying that the ground state is degenerate.  This
makes sense since for zero interaction the ground states are
$$ \rb{0}_b \otimes \rb{0}_d \ \ {\rm and\ \ } 
\rb{0}_b\otimes\rb{1}_d,\eqno\zp$$
\edef\gsZ{\zen}%
where $\rb{0}_b$ is the quasi-particle ground state.  In terms of the 
quasi-electron and quasi-hole operators $(e(k),h(-k))$ with
$k>0$, $\rb{0}_b$ is a zero energy state satisfying
$$ e(k) \rb{0}_b = h(-k) \rb{0}_b = 0.\eqno\zp $$
As $\Eimp$ increases above zero, the degeneracy is lifted by the interaction.
The states (\gsZ) are split into a ground state and the 
bound state  with energy (\eeZ).
In the Ising picture the degenerate ground states, $\rb{0,\pm}_I$, 
are labeled by the 
expectation values of the spin field $\s(x)$, 
$\langle\s(x)\rangle_\pm =\pm {\overline \s}$, where ${\overline \s}$ is the 
spontaneous magnetization.
A non-vanishing magnetic field, $\sqrt{\Eimp}>0$,  removes 
the degeneracy of the free boundary spin states $\rb{0,\pm}_I$.

Based on the above discussion, we can interpret Andreev reflection 
of the quasi-particles as follows.
Suppose we extend our system by adjoining a lattice of 
impurities for $x>0$.  In the region $x>0$ superconductivity 
is more or less destroyed and we have a 
superconductor-normal conductor boundary at $x=0$.
In this case there will be Andreev reflection at the boundary, as well as
transmission.  A lattice consisting of a single impurity then behaves as
a superconductor-normal conductor barrier with the probability for
Andreev reflection unity, $|R_b|^2 = 1$,
just as would be the case for subgap particles
incident on a superconductor.

So far we have only been dealing with one of the Hamiltonians $H_j$ in
(\hsplitZ).  Each type of quasi-particle, 
$b_1^\dg(k)$ or $b_2^\dg(k)$, will form a bound state.  If we represent the
ground state of $H_j$ by $\rb{0}_j$ (for small $\Eimp>0$) and the bound state 
by $\rb{b}_j$, then the ground state of $H= H_1 + H_2$, $\rb{0}$, will be
$$ \rb{0} = \rb{0}_1\otimes\rb{0}_2, \eqno\zp $$
and the excited states are
$$ \rb{1} = \rb{b}_1\otimes \rb{0}_2,\quad\quad 
\rb{1^\prime} = \rb{0}_1\otimes\rb{b}_2, \eqno\zp $$
both with the same energy and lying inside the gap.

\zmsecspc
\noindent
{\zmsec \zsecfour. Conclusions}\par
\nobreak
\zmtsecspc

We have shown that the model (\horigZ) is a
solvable system, which is equivalent to two decoupled
Ising models with boundary magnetic fields.  {}From the boundary 
field theory we calculated the
reflection amplitudes for the quasi-particles.  
At the boundary the quasi-particles are Andreev reflected.
Provided the magnetic energy $\Eimp$ (or equivalently the
hybridization matrix element/magnetic field $V$) is not too large,
bound states exist, consisting of a quasi-particle bound to the impurity.
The energy of the stable bound states is less than the gap,
$E_b < \dd$.  

We hope this calculation further motivates the use of 
techniques common to integrable systems for the study of impurities in
superconductors.  It would be interesting to apply the techniques 
discussed here, and in [\bassilec],  to more complicated systems, 
such as the original Anderson model, or 
 magnetic 
impurities in $d$-wave superconductors studied in [\fradkin].

\zmsecspc
\noindent
{\zmsec Acknowledgments}\par
\nobreak
\zmtsecspc

We would like to thank V. Ambegaokar, A. Clerk and E. Fradkin for 
discussions.  This work is supported in part by the National Science
Foundation and the National Young Investigator program.
\ \   Z.\ S.\ Bassi also acknowledges support from the Olin Foundation.

\zmsecspc

\def\ijmpa{Int.\ J.\ Mod.\ Phys.\ A\ }
\def\jltp{J.\ Low Temp.\ Phys.\ }

\def\jpsj{J.\ Phys.\ Soc.\ Jpn.\ }
\def\np{Nucl.\ Phys.\ }
\def\pr{Phys.\ Rev.\ }

\def\prb{Phys.\ Rev.\ B\ }

\def\pla{Phys.\ Lett.\ A\ }

\def\prl{Phys.\ Rev.\ Lett.\ }
\def\ptp{Prog.\ Theor.\ Phys.\ }

\def\spj{Sov.\ Phys.\ JETP\ } 
\def\spjr{Zh.\ Eksp.\ Teor.\ Fiz.\ }

\def\zp{Z.\ Phys.\ }

\newdimen\zbibindent
\zbibindent=25pt
\def\box#1{\par\vskip5pt\noindent
\hangindent=\zbibindent
\hbox to \zbibindent{{\tt [#1]}\hfil}%
\ignorespaces}

\noindent
{\zmsec References}

\zmtsecspc

\box{1}A. A. Abrikosov and L. P. Gor'kov, \spj {\bf 12} (1961) 1243
[\spjr {\bf 39} (1960) 1781].

\box{2}J. Kondo, \ptp {\bf 32} (1964) 37.

\box{3}P. W. Anderson, \pr {\bf 124} (1961) 41.

\box{4}S. V. Vonsovsky, Yu A. Izyumov and E. Z. Kurmaev, 
{\it Superconductivity of Transition Metals}, Springer-Verlag, 1982.

\box{5}E. M{\"u}ller-Hartmann and J. Zittartz, \zp {\bf 232} (1970) 11;
{\it ibid}. {\bf 234} (1970) 58; J. Zittartz, \zp {\bf 237} (1970) 419;
E. M{\"u}ller-Hartmann, ``Recent Theoretical Work on Magnetic Impurities in
Superconductors'' in {\it Magnetism, Vol.\ 5}, Ed.\ H. Suhl, 
Academic Press, 1973.

\box{6}K. Satori, H. Shiba, O. Sakai and Y. Shimizu, \jpsj {\bf 61}
(1992) 3239; \hfil\break {\it ibid}. {\bf 62} (1993) 3181.

\box{7}M. Jarrell, D. S. Sivia and B. Patton, \prb {\bf 42} (1990) 4804;
W. Chung and M. Jarrell, \prl {\bf 77} (1996) 3621.

\box{8}L. S. Borkowski and P. J. Hirschfeld, \jltp {\bf 96} (1994) 185.

\box{9}V. I. Rupasov, \pla {\bf 237} (1997) 80.

\box{10}A. A. Zvyagin and P. Schlottmann, \prb {\bf 56} (1997) 300.

\box{11}A. Yazdani, B. A. Jones, C. P. Lutz, M. F. Crommie and D. M. Eigler,
Science {\bf 275} (1997) 1767.

\box{12}Z. S. Bassi and A. LeClair, \np B {\bf 552} (1999) 643. 

\box{13}V. M. Filyov and P. B. Wiegmann, \pla {\bf 76} (1980) 283.

\box{14}K. Machida and F. Shibata, \ptp {\bf 47} (1972) 1817.

\box{15}M. Tinkham, {\it Introduction to Superconductivity}, 
McGraw-Hill, Inc., 1996.

\box{16}S. Ghoshal and A. Zamolodchikov, \ijmpa {\bf 9} (1994) 3841.

\box{17}C. R. Cassanello and E. Fradkin, \prb {\bf 56} (1997) 11246.

\end